\documentclass{elsart}
\usepackage{graphicx,amssymb}
\journal{Physics Letters B}
\begin{document}
\addtolength{\topmargin}{+50pt}

\begin{frontmatter}

\title{Particle-Antiparticle Metamorphosis of Massive
Majorana Neutrinos and Gauginos }

\author{D. V. Ahluwalia-Khalilova}
\ead{d.v.ahluwalia-khalilova@heritage.reduaz.mx}

\address{Theoretical Physics Group,
Fac. de Fisica de la UAZ, Zacatecas, Ap. Postal C-600,
ZAC 98062, Mexico.\\ Present address: 
Department of Mathematics, University
of Zacatecas, Zacatecas, ZAC 98060, Mexico}

\begin{abstract}
Recent results on neutrinoless double beta decay,
as reported by Klapdor-Klein\-groth\-aus {\em et al.,\/}
take us for the first time into the realm of Majorana
spacetime structure. However, this structure has either been 
treated as an afterthought to the Dirac construct; or, when
it has been attended to in its own right, its
 physical and mathematical content was never fully unearthed.
In this {\em Letter,\/} we undertake to remedy the existing situation.
We present a detailed formalism required for the description
of the non-trivial spacetime structure underlying the 
$\nu \rightleftharpoons  \overline{\nu}$ metamorphosis 
--- where $\nu$ generically represents a massive Majorana neutrino, 
or a massive gaugino. 


\end{abstract}
\end{frontmatter}

\def\beq{\begin{eqnarray}}
\def\eeq{\end{eqnarray}}


\def\s{\mbox{\boldmath$\displaystyle\mathbf{\sigma}$}}
\def\J{\mbox{\boldmath$\displaystyle\mathbf{J}$}}
\def\K{\mbox{\boldmath$\displaystyle\mathbf{K}$}}
\def\P{\mbox{\boldmath$\displaystyle\mathbf{P}$}}
\def\p{\mbox{\boldmath$\displaystyle\mathbf{p}$}}
\def\hp{\mbox{\boldmath$\displaystyle\mathbf{\widehat{\p}}$}}
\def\x{\mbox{\boldmath$\displaystyle\mathbf{x}$}}
\def\0{\mbox{\boldmath$\displaystyle\mathbf{0}$}}
\def\bv{\mbox{\boldmath$\displaystyle\mathbf{\varphi}$}}
\def\hbv{\mbox{\boldmath$\displaystyle\mathbf{\widehat\varphi}$}}

\def\bn{\mbox{\boldmath$\displaystyle\mathbf{\nabla}$}}

\def\bl{\mbox{\boldmath$\displaystyle\mathbf{\lambda}$}}
\def\bl{\mbox{\boldmath$\displaystyle\mathbf{\lambda}$}}
\def\br{\mbox{\boldmath$\displaystyle\mathbf{\rho}$}}
\def\1{1}
\def\bfhh{\mbox{\boldmath$\displaystyle\mathbf{(1/2,0)\oplus(0,1/2)}\,\,$}}

\def\mn{\mbox{\boldmath$\displaystyle\mathbf{\nu}$}}
\def\amn{\mbox{\boldmath$\displaystyle\mathbf{\overline{\nu}}$}}

\def\mne{\mbox{\boldmath$\displaystyle\mathbf{\nu_e}$}}
\def\amne{\mbox{\boldmath$\displaystyle\mathbf{\overline{\nu}_e}$}}
\def\rlh{\mbox{\boldmath$\displaystyle\mathbf{\rightleftharpoons}$}}

\def\wm{\mbox{\boldmath$\displaystyle\mathbf{W^-}$}}
\def\hh{\mbox{\boldmath$\displaystyle\mathbf{(1/2,1/2)}$}}
\def\h00h{\mbox{\boldmath$\displaystyle\mathbf{(1/2,0)\oplus(0,1/2)}$}}
\def\znbb{\mbox{\boldmath$\displaystyle\mathbf{0\nu \beta\beta}$}}



\section{Introduction}

Spacetime structures in supersymmetric theories find themselves  
deeply intertwined with the Majorana realization 
of fermionic representation spaces. 

In its simplest form, the algebra of supersymmetry is
generated by fourteen operators. These consist of ten generators 
of the Poincar\'e group, and four anticommuting generators. 
The latter are called 
``supertranslations,'' and are components of a Majorana spinor 
\cite{js}. So, it comes about that 
spacetime structure of supersymmetry finds itself deeply
connected with the Majorana aspect. 

In this context, recent results  of Klapdor-Kleingrothaus 
{\em et al.,\/} \cite{Klapdor1,Klapdor2,ew,a},
which provided a first direct evidence for  neutrinoless double 
beta decay, $0\nu\beta\beta$, are of particular interest.

In its simplest interpretation, the  $0\nu\beta\beta$  experimental signal
arises from the Majorana nature of massive $\nu_e$. 
Even though the experiment by itself does not necessarily 
require a supersymmetric framework for its explanation, 
the indication towards a 
Majorana spacetime structure suddenly acquires a  pivotal importance.

In this {\em Letter\/} we, therefore, undertake an {\em ab initio\/}
look at the spacetime structure of the massive Majorana particles --- whether
they be neutrinos or gaugions. In particular, we focus our attention
on the $\nu \rightleftharpoons  \overline{\nu}$ metamorphosis.

In the next section, we present 
a critique of the conventional Majorana construct in the 
context of  $0\nu\beta\beta$. The critique also provides 
additional motivation for undertaking the reported work.
In Sec. 3, we present an {\em ab initio\/} formulation  Majorana spinors.
In Sec. 4, we examine to what extent it is possible
to associate them with negative and positive energies.
There we find that the particle/antiparticle interpretation of self/antiself
conjugate Majorana spinors fails, thus opening the door to the
$\nu\rightleftharpoons \bar \nu $ metamorphosis leading to
$0\nu2\beta $ decay.
In Sec. 5, we show the insensitivity of Majorana self/antiself conjugate
spinors to the arrow of time.
The explicit construct of the particle-antiparticle
metamorphosis is presented in Sec. 6 in terms of
the momentum-space Feynman propagator.
The {\em Letter\/} closes with 
some brief remarks and a summary.


\section{A critique of conventional Majorana construct in \znbb}

As already noted, in the simplest scenario, the recent results of  
Klapdor-Kleingrothaus {\em et al.\/} are interpreted 
as arising from the Majorana nature of
neutrinos. The essential physics of interest 
lies in the following. A virtual $W^-$, which
induces a weak $n\rightarrow p$ transition, decays   into 
$e^-$ and $\overline{\nu}_e$,  
\beq
W^- \rightarrow e^- + \overline\nu_e\,\, ,\label{w1}
\eeq 
and the emitted $\overline{\nu}_e$
is reabsorbed as an $\nu_e$,
\beq
\nu_e \,\, + W^- \rightarrow e^-\label{w2}\,,
\eeq
where the second (virtual) $W^-$ takes responsibility
for the second $n\rightarrow p$ trans\-forma\-tion.
The net result is the $0\nu\beta\beta$,
\beq
A(N,Z) \rightarrow A(N-2,Z+2) + 2 \,e^-\,,
\eeq 
where $A(N,Z)$ represents
a nucleus carrying $N$ neutrons, and $Z$ protons. 
Now, the  Majorana  nature of $\nu_e$ 
does not mean  ($\times$) that one can naively identify the 
$\overline{\nu}_e$ of (\ref{w1}) with the 
$\nu_e$ of (\ref{w2}), 
\beq
\times:\quad \overline{\nu}_e=\nu_e.
\eeq
This is so
because a neutrino carries a negative helicity in the $m_{\nu_e}/E_{\nu_e}
\rightarrow 0$
limit, and transforms as a $(0,1/2)$ object; while an antineutrino carries
a  positive helicity in the $m_{\nu_e}/E_{\nu_e}
\rightarrow 0$ limit, and transforms as a
$(1/2,0)$ object.
As such there is a $m_{\nu_e}/E_{\nu_e}$ suppressed 
amplitude for the above identification \cite{Furry,RabiM}.
This result arises in a framework in which 
the Majorana field \cite{em1937} 
is obtained from the Dirac field,
\beq
\psi(x) = \int \frac{d^3 p}{(2\pi)^3} \frac{m}{2 p_0}
\sum_{h={\uparrow,\downarrow}}
\left[a_h(\p)\, u_h(\p)\, e^{-i p\cdot x}
+ 
b^\dagger_h(\p)\, v_h(\p) \,e^{+i p\cdot x}
\right]\,,\label{Majorana}
\eeq
through the identification,
\beq
b^\dagger_h(\p)= \zeta\, a^\dagger_h(\p)\,,\label{Fock}
\eeq
where $\zeta$ is a ``creation phase factor'' \cite{bk}.

As is apparent, the above Fock space construct ceases to place fundamentally
{\em neutral\/} fields at the same footing as the intrinsically 
{\em charged\/} fields. Instead, it patches together 
the old Dirac  spacetime structure
of the  $(1/2,0)\oplus(0,1/2)$ representation space,
$ u_h(\p)$ and  $v_h(\p)$, with a new Fock-space constraint as
suggested by Majorana (\ref{Fock}).
In other words, the Fock space is treated correctly, via
Eq. (\ref{Fock}), as is required for a fundamentally neutral field.
However, the underlying representation
space is still obliged to the Dirac spinors. 
The Majorana construct, as used
by the high-energy physics  community,
is a {\em hybrid\/} and it asks for the formulation of a
framework that is of Majorana type both in spinor- and Fock- spaces.

The question  now is, whether it is possible to create  
a fully symmetric framework for the intrinsically neutral fields
that carries an explicit counterpart of Eq. (\ref{Fock}) for its
$(1/2,0)\oplus(0,1/2)$ spinorial structure. In this {\em Letter\/} 
we provide such a formulation. In the process we shall arrive at a 
deeper understanding of the spacetime structure underlying the
 $\nu \rightleftharpoons  \overline{\nu}$ metamorphosis.

The presented formulation is self contained. It carries its
interest beyond understanding $0\nu\beta\beta$, in terms of Majorana
neutrinos, to theorists and phenomenologists interested in left-right
symmetric theories  \cite{RabiM} and  gauginos.

\section{Ab initio formulation of Majorana spinors }

The boost generator for the
$(1/2,0)$ representation space is $-i\,\s/2$, 
and  that for $(0,1/2)$  is  $+i\,\s/2$. Consequently, 
the respective boosts are
\beq
B_R&=&\exp\left(+\, 
\frac{\s}{2}\cdot\bv\right)= \sqrt{\frac{E+m}{2\,m}}
\left(\1_2+\frac{\s\cdot\p}{E+m}\right)
\,,\label{br}\\
B_L&=&\exp\left(-\, 
\frac{\s}{2}\cdot\bv\right)= \sqrt{\frac{E+m}{2\,m}}
\left(\1_2-\frac{\s\cdot\p}{E+m}\right)\,,\label{bl}
\eeq
where the boost parameter is defined as:
\beq
\cosh(\varphi)=\frac{E}{m},\quad
\sinh(\varphi)=\frac{\vert\p\vert}{m},\quad 
{\hbv}
=\frac{\p}{\vert \p\vert}\,.
\eeq
The boosts take a particle at rest  to a particle
moving with momentum $\p$ in the ``boosted frame.'' 
We use the notation in which 
$\1_n$ and $0_n$ represent
$n\times n$ identity and null matrices, respectively.

These well-known results allow for the observation: 
\beq
B_{L,R}^{-1}= B_{R,L}^\dagger \, .
\eeq
Further, if  $\Theta$ is the Wigner's spin-$1/2$ time reversal 
operator,\footnote{
We refrain from  identifying $\Theta$ with $-\,i\, \sigma_2$ because
such identification does not exist for higher-spin $(j,0)\oplus(0,j)$ 
representation spaces. The existence of  Wigner time reversal
operator for all $j$, allows, for fermionic $j$'s,
the introduction of $(j,0)\oplus (0,j)$ Majorana objects.}
\beq
\Theta=
\left(
\begin{array}{cc}
0 & -1 \\
1 & 0
\end{array}
\right)\,, 
\eeq
then 
\beq
\Theta \left[\s/2\right] \Theta^{-1} = -\, \left[\s/2\right]^\ast\,,  
\label{wigner}
\eeq

The above observations imply that: 

\begin{enumerate}

\item
If $\phi_L(\p)$ transforms as a left handed spinor, then
$\left(\zeta_\lambda \Theta\right) \,\phi_L^\ast(\p)$
transforms as a right handed spinor -- where, $\zeta_\lambda$ is
an unspecified phase.

\item
If $\phi_R(\p)$ transforms as a right handed spinor, then
$\left(\zeta_\rho \Theta\right)^\ast \,\phi_R^\ast(\p)$
transforms as a left handed spinor -- where, $\zeta_\rho$ is
an unspecified phase.
\end{enumerate}

As a consequence, the following spinors 
(in Weyl basis) belong to the $(1/2,0)\oplus(0,1/2)$
representation space :
\beq
\lambda(\p) =
\left(
\begin{array}{c}
\left(\zeta_\lambda \Theta\right) \,\phi_L^\ast(\p)\\
\phi_L(\p)
\end{array}
\right)\,,\quad
\rho(\p)=
\left(
\begin{array}{c}
\phi_R(\p)\\
\left(\zeta_\rho \Theta\right)^\ast \,\phi_R^\ast(\p)
\end{array}
\right)\,.
\eeq

The operation of charge conjugation is,
\beq
{C} = 
\left(
\begin{array}{cc}
0_2 & i\,\Theta \\
-i\,\Theta & 0_2
\end{array}
\right) {K}\,,
\eeq
where the operator $K$ complex conjugates any object that appears 
on its right.
Demanding $\lambda(\p )$ and $\rho (\p )$ to 
be self/anti-self conjugate under  $C$, 
\beq
{C} \lambda(\p) = \pm  \lambda(\p)\,,\quad
 {C} \rho(\p) = \pm  \rho(\p)\,,
\eeq
restricts the phases, $\zeta_\lambda$ and
$\zeta_\rho$, to two values:
\beq
\zeta_\lambda= \pm\,i\,,\quad \zeta_\rho=\pm\,i\,.
\eeq
The plus sign in the above equation
yields self conjugate, $\lambda^S(\p)$ and $\rho^S(\p)$ 
spinors; while the minus
sign results in the anti-self conjugate spinors,  $\lambda^A(\p)$ and
$\rho^A(\p)$. 

\subsection{The $\lambda(\p)$ spinors}

To obtain explicit expressions for $\lambda (\p )$, we first write down 
the rest spinors. These are:
\beq
\lambda^S(\0) = 
\left(
\begin{array}{c}
+\,i \,\Theta \,\phi_L^\ast(\0)\\
\phi_L(\0)
\end{array}
\right)\,,\quad
\lambda^A(\0) = 
\left(
\begin{array}{c}
-\,i \,\Theta \,\phi_L^\ast(\0)\\
\phi_L(\0)\, 
\end{array}
\right)\, .
\eeq
Next, we choose the $\phi_L(\0)$ to be helicity eigenstates,
\beq
\s\cdot{\hp} \,\phi_L^\pm (\0)= \pm\,\phi_L^\pm(\0)\,,
\label{x}
\eeq  
and concurrently note that\footnote{{\em Derivation of Eq. (\ref{y}):\/}
Complex conjugating Eq. (\ref{x})
gives, 
\[
{\s}^\ast\cdot{{\hp}} \,\left[\phi_L^\pm (\0)\right]^\ast= 
\pm\,\left[\phi_L^\pm(\0)\right]^\ast\,.  
\]
Substituting for $\s^\ast$ from Eq. (\ref{wigner}) then results in,
\[
\Theta \s \Theta^{-1}\cdot{\hp} \,\left[\phi_L^\pm (\0)\right]^\ast
= 
\mp\,\left[\phi_L^\pm(\0)\right]^\ast\, .
\]
But $\Theta^{-1} = -\Theta$. So, 
\[
- \Theta \s \Theta
\cdot{\hp} \,\left[\phi_L^\pm (\0)\right]^\ast= 
\mp\,\left[\phi_L^\pm(\0)\right]^\ast \,.
\]
Or, equivalently,
\[
\Theta^{-1} \s \Theta
\cdot{\hp} \,\left[\phi_L^\pm (\0)\right]^\ast= 
\mp\,\left[\phi_L^\pm(\0)\right]^\ast \,.
\]
Finally, left multiplying  both sides of the preceding 
equation by $\Theta$, and moving $\Theta$ through 
${\hp}$, yields Eq. (\ref{y}).}
\beq
\s\cdot\hp \, \Theta \left[\phi_L^\pm (\0)\right]^\ast
= \mp\, \Theta\left[\phi_L^\pm(\0)\right]^\ast\,.
\label{y}
\eeq 
We are thus led to four rest spinors. Two of which are self-conjugate, 
\beq
\lambda_\uparrow^S(\0) = 
\left(
\begin{array}{c}
+\,i \,\Theta \,\left[\phi^+_L(\0)\right]^\ast\\
\phi^+_L(\0)
\end{array}
\right)\,,\quad
\lambda_\downarrow^S(\0) = 
\left(
\begin{array}{c}
+\,i \,\Theta \,\left[\phi^-_L(\0)\right]^\ast\\
\phi^-_L(\0)\, 
\end{array}
\right)\, , \quad
\eeq
and the other two, which are anti-self conjugate,
\beq
\lambda_\uparrow^A(\0) = 
\left(
\begin{array}{c}
-\,i \,\Theta \,\left[\phi^+_L(\0)\right]^\ast\\
\phi^+_L(\0)
\end{array}
\right)\,,\quad
\lambda_\downarrow^A(\0) = 
\left(
\begin{array}{c}
-\,i \,\Theta \,\left[\phi^-_L(\0)\right]^\ast\\
\phi^-_L(\0)\, 
\end{array}
\right)\, .
\eeq
The boosted spinors are now obtained via the operation:
\beq
\lambda(\p)=\left(
\begin{array}{cc}
B_R & 0_2 \\
0_2 & B_L
\end{array}
\right)\lambda(\0)\,,\label{z}
\eeq
where the $B_R$ and $B_L$ are given by Eqs. (\ref{br}) and (\ref{bl}).
In the boost, we replace $\s\cdot\p$ by $\s\cdot{\hp}\,\vert \p\vert$,
and then exploit Eq. (\ref{y}). After simplification,
Eq. (\ref{z}) yields:
\beq
\lambda_\uparrow^S(\p)
=
\sqrt{\frac{E+m}{2\,m}}\left(1-\frac{ \vert \p \vert}{E+m}\right)
\lambda_\uparrow^S(\0)\,,\label{lsup}
\eeq
which, in the massless limit, {\em identically vanishes\/}, while
\beq
\lambda_\downarrow^S(\p)
=
\sqrt{\frac{E+m}{2\,m}}\left(1+\frac{ \vert \p \vert}{E+m}\right)
\lambda_\downarrow^S(\0)\, ,
\label{lsdown}
\eeq
does not.
We hasten to warn the reader that one should not be tempted to read the
two different prefactors to $\lambda^S(\0)$
in the above expressions as the boost operator that appears in Eq. 
(\ref{z}). For one thing, there is only one (not two) 
boost operator(s) in the
$(1/2,0)\oplus(0,1/2)$ representation space. The simplification that
appears here is due to a fine interplay between Eq. (\ref{y}), the boost
operator, and the structure of the $\lambda^S(\0)$.
Similarly, the anti-self conjugate set of the boosted spinors reads:
\beq
\lambda_\uparrow^A(\p)
=
\sqrt{\frac{E+m}{2\,m}}\left(1-\frac{ \vert \p \vert}{E+m}\right)
\lambda_\uparrow^A(\0)\,,\label{laup}\\
\lambda_\downarrow^A(\p)
=
\sqrt{\frac{E+m}{2\,m}}\left(1+\frac{ \vert \p \vert}{E+m}\right)
\lambda_\downarrow^A(\0)\,.\label{ladown}
\eeq
In the massless limit, the first of these spinors 
{\em identically vanishes\/}, while the second does not.
Representing the unit vector along $\p$, 
as, 
\beq
\hp =\Big(\sin(\theta)\cos(\phi),\,
\sin(\theta)\sin(\phi),\,\cos(\theta)\Big)\,,
\eeq
the $\phi^\pm_L(\0)$ take the explicit form:
\beq
\phi_L^+(\0) =
\sqrt{m} e^{i\vartheta_1} 
\left(
\begin{array}{c}
\cos(\theta/2) e^{-i\phi/2}\\
\sin(\theta/2) e^{i\phi/2}
\end{array}
\right)\,,\\
\phi_L^-(\0) =
\sqrt{m} e^{i\vartheta_2} 
\left(
\begin{array}{c}
\sin(\theta/2) e^{-i\phi/2}\\
-\cos(\theta/2) e^{i\phi/2}
\end{array}
\right)\,.
\eeq
On setting $\vartheta_1$ and $\vartheta_2$ to be zero --- a fact
that we explicitly note --- we find the following {\em 
bi-orthonormality\/} relations for the self-conjugate Majorana 
spinors,
\beq
&& \overline{\lambda}^S_\uparrow(\p) \lambda^S_\uparrow (\p) = 0\,,
\quad 
\overline{\lambda}^S_\uparrow(\p) \lambda^S_\downarrow (\p) = + 2 i m
\,,\\
&& \overline{\lambda}^S_\downarrow(\p) \lambda^S_\uparrow (\p) = - 2 i m\,,
\quad 
\overline{\lambda}^S_\downarrow(\p) \lambda^S_\downarrow (\p) = 0
\,.
\eeq
Their counterpart for antiself-conjugate Majorana spinors reads, 
\beq
&& \overline{\lambda}^A_\uparrow(\p) \lambda^A_\uparrow (\p) = 0\,,
\quad 
\overline{\lambda}^A_\uparrow(\p) \lambda^A_\downarrow (\p) = - 2 i m
\,,\\
&& \overline{\lambda}^A_\downarrow(\p) \lambda^A_\uparrow (\p) = + 2 i m\,,
\quad 
\overline{\lambda}^A_\downarrow(\p) \lambda^A_\downarrow (\p) = 0
\,,
\eeq
while all combinations  of the type 
$\overline{\lambda}^A(\p) \lambda^S(\p)$ and
$\overline{\lambda}^S(\p) \lambda^A(\p)$ identically vanish.

We take note that the bi-orthogonal norms of the Majorana spinors
are intrinsically {\em imaginary.}
The associated completeness relation is:

\beq
-\frac{1}{2 i m}
\left(\left[\lambda^S_\uparrow(\p) \overline{\lambda}^S_\downarrow(\p)
-\lambda^S_\downarrow(\p) \overline{\lambda}^S_\uparrow(p)\right] 
-
\left[\lambda^A_\uparrow(\p) \overline{\lambda}^A_\downarrow(\p)
-\lambda^A_\downarrow (\p)\overline{\lambda}^A_\uparrow(\p)\right]\right)
 = \1_4\,.\label{lc}\nonumber\\
\eeq

\subsection{The $\lambda(\p)$ and $\rho(\p)$ spinors}

Now, $(1/2,0)\oplus(0,1/2)$ is a four dimensional representation space.
Therefore, there cannot be more than four independent spinors.
Consistent with this observation, we find that 
the $\rho(\p)$ spinors are related to the $\lambda(\p)$ spinors 
via the following identities:
\beq
&&\rho^S_\uparrow(\p) = - i \lambda^A_\downarrow(\p)\,,\quad
\rho^S_\downarrow(\p) = + i \lambda^A_\uparrow(\p),\label{id1}\\
&&\rho^A_\uparrow(\p) = + i \lambda^S_\downarrow(\p)\,,\quad
\rho^A_\downarrow(\p) = - i \lambda^S_\uparrow(\p)\,.\label{id2}
\eeq
Using these identities, one may immediately obtain the bi-orthonormality
and completeness relations for the $\rho(\p)$ spinors.

In the massless limit, $\rho^S_\downarrow(\p)$ and $\rho^A_\downarrow(\p)$
{\em identically vanish.\/}

A particularly simple orthonormality, as opposed to bi-orthonormality,
relation exists between the 
$\lambda(\p)$ and $\rho(\p)$ spinors:
\beq
\overline{\lambda}^S_\uparrow(\p) 
\rho^A_\uparrow(\p) = -2 m = \overline{\lambda}^A_\uparrow(\p)
\rho^S_\uparrow (\p)\\  
\overline{\lambda}^S_\downarrow(\p) 
\rho^A_\downarrow(\p) = + 2 m = \overline{\lambda}^A_\downarrow(\p)
\rho^S_\downarrow (\p).
\eeq
An associated completeness relation also exists, and it reads:
\beq
-\frac{1}{2  m}
\left(\left[\lambda^S_\uparrow(\p) \overline{\rho}^A_\uparrow(\p)
+\lambda^S_\downarrow(\p) \overline{\rho}^A_\downarrow(p)\right] 
+
\left[\lambda^A_\uparrow(\p) \overline{\rho}^S_\uparrow(\p)
+\lambda^A_\downarrow (\p)\overline{\rho}^S_\downarrow(\p)\right]\right) 
= \1_4\,.\nonumber\\
\label{lrcompleteness}
\eeq
The results of this section are in spirit of Refs.~\cite{Mc,Case,Ah2,Ah3}.

\subsection{Reflection-less Majorana spinors}

In the massless limit, the Majorana spinors are reflection-less.
This can be proved by taking note of the fact that, e.g., 
under the Parity operation $\lambda_{\downarrow}^S(\p)$ transforms to
$ i \lambda^S_\uparrow (- \p)$. The latter, in the indicated limit, 
identically vanishes, as already explained in subsection (3.1). 
As a consequence, for instance, 
in the ordinary beta decay
\beq
n\rightarrow p + e^- + \overline{\nu}_e\,,
\eeq
if one identifies $\overline{\nu}_e$ with $\rho^A_\uparrow(\p)$ [which equals
$i \lambda_\downarrow^S(\p)$,  according to the first
equation in Eq. (\ref{id2})],
then its Parity transform $-\lambda^S_\uparrow(-\p)$,
identically vanishes in the massless limit.
This picture is in accord with the empirical observation that 
the origin of parity violation in ordinary beta decay lies in the
absence of its mirror image. 
For massive neutrinos, this would  be no longer true. 

Note a rather subtle aspect of the Majorana framework: 
Parity violation occurs despite the apparent parity covariance of 
the Majorana construct.\footnote{   
This circumstance is akin to the violation of special relativity
in Rarita-Schwinger framework despite its
fully  Poincar\'e covariant structure (see, Refs. 
\cite{vz,ka}).} No such kinematic conclusion can be drawn 
if neutrinos are described by Dirac spinors without explicitly
invoking parity violation via the standard projectors 
$(1/2)(1\pm\gamma_5)$.

\subsection{Wave equations for Majorana spinors}

To analyze the time-evolution of the $\lambda(\p)$ and $\rho(\p)$ Majorana
spinors one needs relevant equations of motion. After 
taking into account the observations of Ref. \cite{Ah4,gga},
these can be obtained   
following a text-book procedure similar to that of Ryder \cite{lhr}. 
The result is:
\beq
\left(
\begin{array}{cc}
-\, \1_2 &   {\mathcal O}^+_\lambda \\
 {\mathcal O}^-_\lambda  & -\,\1_2
\end{array}
\right)\lambda(\p) =0\,,\quad
\left(
\begin{array}{cc}
+\, \1_2 & {\mathcal O}^+_\rho \\
{\mathcal O}^-_\rho & +\,\1_2
\end{array}
\right)\rho(\p) =0\,.\label{weq}
\eeq
In the above equations we have used the following abbreviations:
\beq
{\mathcal O}^\pm_\lambda&\equiv&
 \zeta_\lambda \exp\left(\pm\s\cdot\bv/2\right)
\Theta\eta\exp\left(\pm\s\cdot\bv/2\right)\,,\\
{\mathcal O}^\pm_\rho&\equiv&
 \zeta_\rho \exp\left(\pm\s\cdot\bv/2\right)
\Theta\eta\exp\left(\pm\s\cdot\bv/2\right)\,,
\eeq
where
\beq
\eta=\left(
\begin{array}{cc}
e^{i \phi} & 0 \\
0 & e^{-i \phi}
\end{array}
\right)\,.
\eeq
These wave equations {\em cannot\/} be reduced to the Dirac 
equation. 

\section{Indiscernibility  between
negative and positive energy solution for Majorana particles }

We note that for self-, as well antiself- conjugate Majorana 
spinors (i.e., irrespective of whether $\zeta_\lambda=+i$, or
 $\zeta_\lambda=-i$) 
\beq
\mbox{Det}\left(
\begin{array}{cc} 
-\, \1_2 &   {\mathcal O}^+_\lambda \\
 {\mathcal O}^-_\lambda  & -\,\1_2
\end{array}
\right) =\left(3 m^2 + 4E m-\p^2 + E^2 \right)^2 
\left(m^2+\p^2-E^2\right)^2\,.
\label{det}
\eeq
That the determinant must vanish for  non-trivial solutions, $\lambda(\p)$,
translates into,  $\left(m^2+\p^2-E^2\right)=0$.
The $\lambda(\p)$ spinors are thus consistent
with the  causal dispersion relation.
To check the hypothesis if the sign of $\zeta_\lambda$ can determine
the sign of energy associated with the $\lambda(\p)$ spinors, it suffices
to look at the rest spinors.  
The wave equation for the latter, irrespective of the helicity,
or the sign of $\zeta_\lambda$, then reduces to the conditions of the 
type\footnote{Equation below is written for 
$\lambda^S_\uparrow(\0)$.}
\beq
\left(
\begin{array}{c}
i e^{-i \phi/2} \left(m^2-E^2\right) \sin(\theta/2)\\
- i e^{i \phi/2} \left(m^2-E^2\right) \cos(\theta/2)\\
- e^{-i \phi/2} \left(m^2-E^2\right) \cos(\theta/2)\\
- e^{i \phi/2} \left(m^2-E^2\right) \sin(\theta/2)
\end{array}
\right)
=
\left(
\begin{array}{c}
0\\
0\\
0\\
0
\end{array}
\right)\,.
\eeq
Clearly, this condition cannot determine the sign of 
energy associated with the $\lambda^S_\uparrow(\0)$ to be either plus,
or minus.
The same remains true for the remaining $\lambda(\0)$ spinors, and results
readily extend to all $\lambda(\p)$.

The above discussion  is equally valid for the $\rho(\p)$ spinors

\section{Insensitivity of Majorana spinors  to the direction
of time}

The plane waves for the $\lambda(\x,t)$ and $\rho(\x,t)$
are, $\exp(i\, \epsilon\, p\cdot x)\lambda(\p)$ and $
\exp(i\,\epsilon\, p\cdot  x)\rho(\p)$, where $\epsilon=\pm 1$ (depending 
upon whether the propagation is forward in time, or backward in time).
To determine $\epsilon$, it suffices to study the wave equations in the
rest frame of the particles. In that frame, the $\lambda(\x,t)$
satisfies the following (simplified)differential 
equation:\footnote{To obtain the simplified 
equation below we first multiplied 
the momentum-space wave equation by $2m (E+m)$. Then, we 
exploited the fact 
that in configuration space, $\p=\frac{1}{i} {\bn}$. 
When it acts  upon 
$\lambda(\0)$ the resulting eigenvalue 
vanishes (so we dropped this term), 
and that $E=i \frac{\partial}{\partial t}$.} 
\def\a{- 2 m \left(i \frac{\partial}{\partial t} + m\right)}

\def\b{- \zeta_\lambda e^{-i\phi}\left(m^2 - \frac{\partial^2}{\partial t^2}
+ 2 i m \frac{\partial}{\partial t}\right)}

\def\c{\zeta_\lambda e^{i\phi}\left(m^2 - \frac{\partial^2}{\partial t^2}
+ 2 i m \frac{\partial}{\partial t}\right)}

\def\oa{{\mathcal O}_a}
\def\ob{{\mathcal O}_b}
\def\oc{{\mathcal O}_c}
\beq
\left(
\begin{array}{cccc}
\oa & 0  & 0 & \ob  \\
0  & \oa & \oc & 0\\ 
0  & \ob & \oa & 0\\
\oc & 0 & 0 & \oa
\end{array}
\right)\lambda(\0) e^{-i \epsilon\, m t} = 0
\eeq
where
\beq
&&\oa\equiv\a\,,\\
&&\ob\equiv\b\,,\\
&&\oc\equiv\c\,.
\eeq
This equation does not fix the sign of $\epsilon$.
It only determines $\epsilon^2$ to be unity. That is, 
Majorana spinors are insensitive to the {\em forward\/} and
{\em backward\/} directions in time so important in the Feynman-St\"uckelberg
interpretation of particles and antiparticles.
Same result, and similar analysis, holds true for the $\rho$ spinors.
The {\em conventional\/} 
distinction between particles and antiparticles
disappears.  However,  a distinction does reside in the 
behavior of these spinors as regards the high-energy/massless limit and the
associated helicity.
In that limit, as is apparent from Eqs. (\ref{lsup}) and   
(\ref{lsdown}), only the {\em negative\/} helicity $\lambda(\p)$
survive.
Similarly, Eqs. (\ref{laup}) and 
(\ref{ladown}) show that in the same limit
only the {\em positive\/} helicity $\rho(\p)$ survive.
This circumstance suggests that we 
identify the $\lambda(\p)$ as  the {\em particle
spinors;\/} and call the $\rho(\p)$ spinors  
{\em antiparticle
spinors,\/} in accord with the common terminology 
prevalent in neutrino physics. However, 
this identification is very different in spirit from the original notion
of particles and antiparticles.
As such, it appears advised to introduce a new term
that specifies these particles and antiparticles as 
{\em Majorana particles/antiparticles.\/}  

As regards their properties under the charge and charge conjugation 
operators, there are thus two type of particles in Nature. 
The $e^\pm$, $\mu^\pm$, $\tau^\pm$, and the known quarks, 
are all of the first
--- Dirac --- type. They are eigenstates of the charge operator. 
The Klapdor-Kleingrothaus {\em et al.\/}
\cite{Klapdor1,Klapdor2} experiment addresses itself
to  $\nu_e$ and $\overline\nu_e$; and finds them to be of Majorana
type. This second type of particles are eigenstates of the
charge conjugation operator. So beyond pinning down the masses
of  $\nu_e$ and $\overline\nu_e$, discovering lepton number
violation, the  Klapdor-Kleingrothaus {\em et al.\/} 
experiment for the first time
takes us into a realm of spacetime structure which so far had remained
only in the sphere of theoretical constructs. That newly 
unearthed structure, as we are making the case here, must be 
treated in its own right, and not as an afterthought to the
celebrated Dirac construct.

\section{Metamorphosis of Majorana $\amn$ to $\mn $:
Propagator for the Majorana field}

The completeness relation (\ref{lc})
suggests that we define the Majorana field operator
as:
\beq
\mu(x) = \int\frac{d^3\p}{(2\pi)^3} \frac{1}{2 p_0}
\sum_h\left(\lambda_h^S(\p) a_h(\p) e^{-i p\cdot x}
+ \lambda_h^A(\p) a^\dagger_h(\p) e^{i p\cdot x}\right)
\,.
\eeq
Parenthetically, we take note of the fact that,
using Eqs. (\ref{id1}), it may be re-written entirely in terms of
the self-conjugate $\lambda^S(\p)$ and $\rho^S(\p)$ spinors.
Any relative signs that different helicity Majorana spinors
may carry in the indicated sum are implicitly assumed to be absorbed in 
the annihilation, $a_h(\p)$, and creation, $a^\dagger_h(\p)$, 
operators. The latter are assumed to satisfy the 
bi-orthogonality  respecting fermionic anticommutators,
\beq
\left\{
a_h(\p), \,a^\dagger_{h^\prime}(\p^\prime)\right\}
= (2\pi)^3 \,2 p_0\, \delta^3(\p-\p^\prime)\,
\delta_{h,-h^\prime}
\,.\label{anticomm}
\eeq
By postulating this field-theoretic structure we hope to avoid 
negative norm states, while preserving the essential 
bi-orthonormality of the underlying\-\-\-
$(1/2,0)\oplus(0,1/2)$ Majorana representation space.
In equation (\ref{anticomm}), if, $h=\uparrow$, then $-h$ represents
$\downarrow$, and 
if, $h=\downarrow$, then $-h$ represents
$\uparrow$. With these definitions in mind,
the configuration-space Feynman-Dyson (FD) propagator 
is obtained as
\beq
\langle x \vert S_{FD}\vert y\rangle 
= \langle {\mbox{vac}} \vert T\left[\mu(x)\overline\mu(y)\right]
\vert{\mbox{vac}}\rangle\,,
\eeq
in the standard contextual meaning of the used symbols.
A straightforward calculation yields:
\beq
\langle x \vert S_{FD}\vert y\rangle 
=&& \sum_h \int\frac{d^3\p}{(2\pi)^3} \frac{1}{2 p_0} 
{\Big[}
\lambda^S_h(\p) \overline\lambda^S_{\{-h\}}(\p)\, e^{-i p\cdot(x-y)}
\,\theta\left(x^0-y^0\right)\nonumber\\
 && -\; \lambda^A_h(\p) \overline\lambda^A_{\{-h\}}(\p)\, e^{i p\cdot(x-y)}
\,\theta\left(y^0-x^0\right){\Big]}\,.
\eeq
As a consequence, the momentum-space propagator is obtained to be,
\beq
\langle p^\prime \vert S_{FD} \vert p\rangle
&=& \int\frac{d^4 x}{ (2\pi)^2}
    \int\frac{d^4 y}{ (2\pi)^2}
e^{i p \cdot x}
e^{-i p^\prime\cdot y} \langle x \vert S_F\vert y\rangle \nonumber\\
&=& i \delta^4(p^\prime - p) \sum_h
\left(
\frac{\lambda^S_h(\p) \overline\lambda^S_{\{-h\}}(\p)}
{p_0 - E(\p)+ i\epsilon } -
\frac{\lambda^A_h(- \p) \overline\lambda^A_{\{-h\}}(- \p)}
{p_0 + E(\p)- i\epsilon }
\right)\,,\label{prop}
\eeq
where $E(\p)= +\sqrt{m^2+\p^2}$.
As the above propagator explicitly exhibits,
an exchanged Majorana particle is emitted with one helicity,
at one vertex, and is
absorbed with {\em opposite\/} helicity, 
at the second vertex. This is the spacetime picture 
of ${\overline\nu}\rightleftharpoons\nu$ metamorphosis 
for Majorana particles. Using Eqs. (\ref{id1}) and (\ref{id2}), 
the preceding statement can be made more transparent.
Exploiting the indicated identities, 
one may re-express $\langle p^\prime \vert S_{FD} \vert p\rangle$
by replacing the $\overline\lambda^S_{\{-h\}}(\p)$ and 
$\overline\lambda^A_{\{-h\}}(- \p)$ in terms of $\overline\rho^A_{\{h\}}(\p)$
and $\overline\rho^S_{\{h\}}(- \p)$, respectively,
\beq
\langle p^\prime \vert S_{FD} \vert p\rangle
= \delta^4(p^\prime - p)&& \sum_h
{\Bigg(}
\frac{- \lambda^S_\uparrow(\p) \overline\rho^A_\uparrow(\p) 
+ \lambda^S_\downarrow(\p) \overline\rho^A_\downarrow(\p)}
{p_0 - E(\p)+ i\epsilon }   \nonumber \\
&& -
\frac{ \lambda^A_\uparrow(- \p) \overline\rho^S_\uparrow(- \p)
 - \lambda^A_\downarrow(- \p) \overline\rho^S_\downarrow(- \p)
}{p_0 + E(\p)- i\epsilon }
{\Bigg)}\,,\label{prop_rw}
\eeq
In the massless limit, depending on $h$, 
either $\lambda_h(\p)$, or $\rho_h(\p)$, 
identically vanishes. Thus, each of the $\lambda_h(\p)\overline{\rho}_h(\p)$ 
terms   contained in $\langle p^\prime \vert S_{FD} \vert p\rangle$
also vanishes. In the massless limit, this results in a non-existent (i.e.,
identically vanishing) propagator. This implication is in 
accord with the conventional framework which also predicts
that the amplitude for $0\nu\beta\beta$ vanishes for massless
neutrinos. The new propagator, in addition, manifestly shows how the 
$\nu \rightleftharpoons \overline \nu$ metamorphosis 
arises from the underlying spacetime structure.

For comparison, the standard Dirac propagator can be cast \cite{ijmpe} 
into a form similar to (\ref{prop}):   
\beq
\langle p^\prime \vert S_{FD} \vert p\rangle
= i \delta^4(p^\prime - p) \sum_h
\left(
\frac{u_h(\p) \overline{u}_h(\p)}
{p_0 - E(\p)+ i\epsilon } -
\frac{v_h(- \p) \overline v_h(- \p)}
{p_0 + E(\p)- i\epsilon }
\right)\, .\label{prop_dirac}
\eeq

\section{Concluding remarks}

Klapdor-Kleingrothaus {\em et al.\/} experiment
provides first evidence towards the Majorana nature
of massive  $\nu_e$, and 
the associated $\nu_e\rightleftharpoons\overline{\nu}_e$ metamorphosis.
It is an encouraging development towards
discovering new spacetime structures beyond the time-honored
spacetime constructs. Of these supersymmetry is the most natural
and sought after.\footnote{ 
Besides predictions of supersymmetry, discovery of Wigner-type
particles \cite{bww}, where a massive spin one particle 
and its antiparticle carry {\em opposite\/} relative intrinsic parities,
would be a significant benchmark for new type of matter fields
in nature. Equally important would be confirmation
of the new spacetime structures associated with massive
gravitino \cite{ka} and the massive gauge fields \cite{ak}.}

Here, we defined Majorana spinors as self/antiself conjugate eigenspinors
of the charge conjugation operator.  These spinors have null norm; and 
their bi-orthonormal norm is necessarily imaginary. We 
obtained the associated completeness relation, and the relevant 
wave equation. We considered 
the notion of particle and antiparticle for Majorana particles, and
examined the latter aspect in the context of Dirac and  
Feynman-St\"uckelberg  frameworks. Finally, we presented the Majorana 
field as a quantum object, and obtained the Feynman-Dyson propagator. 
The latter explicitly exhibits the   
$\overline{\nu}\rightleftharpoons\nu$ metamorphosis for Majorana
particles, whether they be neutrinos, or massive gaugions.

Figure 1 presents the $\nu_e\rightleftharpoons\overline{\nu}_e$ 
metamorphosis as it emerges from the formalism presented in 
this {\em Letter.\/} In presenting this spacetime picture we 
do not necessarily assume the neutrinos to be left-handed 
objects; thus leaving room for physics beyond the 
standard model. Generalization of this spacetime picture to 
supersymmetric frameworks, and left-right symmetric
theories, is implicit in the presented formalism.

\begin{figure}
\begin{center}
\vskip -3 cm
\includegraphics*[width=11cm]{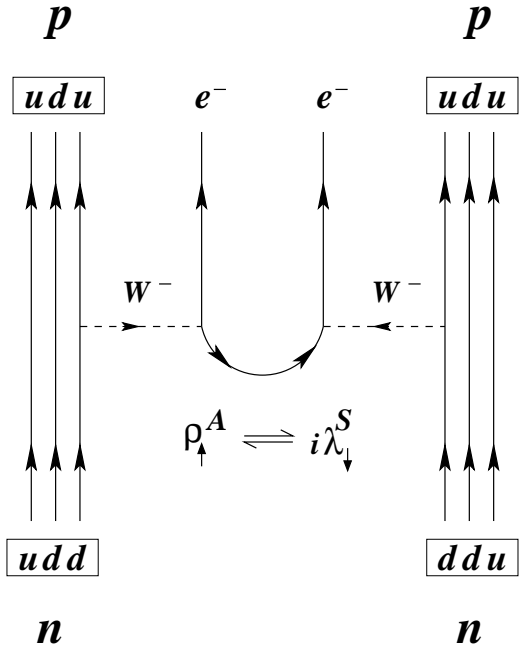}
\end{center}
\caption{ $\nu_e\rightleftharpoons \bar \nu_e $ metamorphosis in
$0\nu\beta \beta $ as seen from the vantage point of the formalism 
presented in this {\em Letter.\/} 
}
\end{figure}

\section*{Acknowledgments}
This work is supported by Consejo Nacional de Ciencia y
Tecnolog\'ia (CONACyT, Mexico) under grant number 32067-E, and by
an affiliate agreement with the Los Alamos National Laboratory.
Thanks are extended to T. Goldman, Y. Liu, and U. Sarkar for 
useful remarks on an earlier draft of this work.

\end{document}